\documentclass{aa}
\usepackage[dvips]{graphicx}

\usepackage{txfonts}
\usepackage[comma,authoryear]{natbib}
\usepackage[latin1]{inputenc}

\usepackage[dvips]{graphicx}

\def\bbbc{{\mathchoice {\setbox0=\hbox{$\displaystyle\rm C$}\hbox{\hbox
to0pt{\kern0.4\wd0\vrule height0.9\ht0\hss}\box0}}
{\setbox0=\hbox{$\textstyle\rm C$}\hbox{\hbox
to0pt{\kern0.4\wd0\vrule height0.9\ht0\hss}\box0}}
{\setbox0=\hbox{$\scriptstyle\rm C$}\hbox{\hbox
to0pt{\kern0.4\wd0\vrule height0.9\ht0\hss}\box0}}
{\setbox0=\hbox{$\scriptscriptstyle\rm C$}\hbox{\hbox
to0pt{\kern0.4\wd0\vrule height0.9\ht0\hss}\box0}}}}
\def\bbbq{{\mathchoice {\setbox0=\hbox{$\displaystyle\rm
Q$}\hbox{\raise
0.15\ht0\hbox to0pt{\kern0.4\wd0\vrule height0.8\ht0\hss}\box0}}
{\setbox0=\hbox{$\textstyle\rm Q$}\hbox{\raise
0.15\ht0\hbox to0pt{\kern0.4\wd0\vrule height0.8\ht0\hss}\box0}}
{\setbox0=\hbox{$\scriptstyle\rm Q$}\hbox{\raise
0.15\ht0\hbox to0pt{\kern0.4\wd0\vrule height0.7\ht0\hss}\box0}}
{\setbox0=\hbox{$\scriptscriptstyle\rm Q$}\hbox{\raise
0.15\ht0\hbox to0pt{\kern0.4\wd0\vrule height0.7\ht0\hss}\box0}}}}
\def\bbbt{{\mathchoice {\setbox0=\hbox{$\displaystyle\rm
T$}\hbox{\hbox to0pt{\kern0.3\wd0\vrule height0.9\ht0\hss}\box0}}
{\setbox0=\hbox{$\textstyle\rm T$}\hbox{\hbox
to0pt{\kern0.3\wd0\vrule height0.9\ht0\hss}\box0}}
{\setbox0=\hbox{$\scriptstyle\rm T$}\hbox{\hbox
to0pt{\kern0.3\wd0\vrule height0.9\ht0\hss}\box0}}
{\setbox0=\hbox{$\scriptscriptstyle\rm T$}\hbox{\hbox
to0pt{\kern0.3\wd0\vrule height0.9\ht0\hss}\box0}}}}
\def\bbbz{{\mathchoice {\hbox{$\sf\textstyle Z\kern-0.4em Z$}}
{\hbox{$\sf\textstyle Z\kern-0.4em Z$}}
{\hbox{$\sf\scriptstyle Z\kern-0.3em Z$}}
{\hbox{$\sf\scriptscriptstyle Z\kern-0.2em Z$}}}}
\newcommand{\beq}{\begin{equation}}
\newcommand{\beqa}{\begin{eqnarray*}}
\newcommand{\beqan}{\begin{eqnarray}}
\newcommand{\greq}{\begin{equation}\left\{ \begin{array}{l}}
\newcommand{\egreq}{\end{array}\right. \end{equation}}
\newcommand{\nngreq}{\[\left\{ \begin{array}{l}}
\newcommand{\nnegreq}{\end{array}\right. \]}
\newcommand{\egreqn}[1]{\end{array}\right. \label{#1}\end{equation}}
\newcommand{\eeq}{\end{equation}} 
\newcommand{\eeqn}[1]{\label{#1}\end{equation}} 
\newcommand{\eeqa}{\end{eqnarray*}}
\newcommand{\eeqan}[1]{\label{#1}\end{eqnarray}}

\newcommand{\bv}{\overline{v}}

\newcommand{\demi}{\frac{1}{2}}

\begin{document}
\title{Solar supergranulation revealed by granule tracking}
\titlerunning{Solar supergranulation}

\author{Michel Rieutord, Nadège Meunier\thanks{\emph{Present address:}
Laboratoire d'Astrophysique, Observatoire de Grenoble, BP 53,
38041 Grenoble cedex 9}, Thierry Roudier, Sylvain Rondi, Francis
Beigbeder and Laurent Parès}

\authorrunning{Rieutord et al.}

\institute{
Laboratoire d'Astrophysique de Toulouse et Tarbes, UMR 5572, CNRS et
Université Paul Sabatier Toulouse 3, 14 avenue E. Belin, 31400 Toulouse,
France\\
\email{\tiny [rieutord,roudier,francis.beigbeder,pares]@ast.obs-mip.fr,
nmeunier@obs.ujf-grenoble.fr, rondi@wanadoo.fr}}
\date{\today}

\offprints{M. Rieutord}

\abstract{% Context
Supergranulation is a pattern of the velocity field at the surface of the
Sun, which has been known about for more than fifty years,
however, no satisfactory explanation of its origin has been proposed.
}{%Aims
New observational constraints are therefore needed to guide theoretical
approaches which hesitate between scenarios that either invoke a
large-scale instability of the surface turbulent convection or a direct forcing
by buoyancy.
}{%Method
Using the 14-Mpixel CALAS camera at the Pic-du-Midi observatory, we obtained
a 7.5~h-long sequence of high resolution images with unprecedented field size.
Tracking granules, we have determined the velocity field
at the Sun's surface in great detail from a scale of 2.5~Mm up to 250~Mm.
}{% Results
The kinetic energy density spectrum shows that supergranulation peaks
at 36~Mm and spans on scales ranging between 20~Mm and 75~Mm. The
decrease of supergranular flows in the small scales is close to
a $k^{-2}$-power law, steeper than the equipartition Kolmogorov one.
The probability distribution function of the divergence field shows
the signature of intermittency of the supergranulation and thus its
turbulent nature.
}{% Conclusions
}

\maketitle

\section{Introduction}

Supergranulation was discovered by \cite{H54} using Doppler
images of the Sun. It appeared as an essentially horizontal flow field
at a typical scale of 30~Mm. The origin of this flow field was first
thought to be related to the second ionization of helium, which provides
some latent heat at a depth around 10~Mm, compatible with their typical
size (\citealt{SW68}). This scenario has been much debated because of
the weakness of the effect and the apparent vigour of the supergranular
flow. Other ways of generating this velocity scale rely on large-scale
instabilities of the surface turbulent flow (\citealt{RRMR00}), triggered
by the strong density stratification.  In this scenario, kinetic energy
of granules, the small-scale convective cells, is thought to be piped
to a larger scale by an AKA-like effect (\citealt{GVF94}). However,
other kinds of large-scale instabilities are still possible, like a convective
instability triggered by fixed-flux boundary conditions imposed by the
small-scale granular convection (\citealt{RR03b}). In this approach, the
cooling resulting from the granulation does not suppress the convective
instability of the larger scales; because all the heat flux is carried
by the small scale, none is carried by the large scale instability, which
shows little temperature fluctuations (see \citealt{MTR07}). Besides these
scenarios, recent observations of \cite{GDS03} added some wavelike
properties to supergranulation.

Obviously, the supergranulation theory needs guidance from every available
observational constraint. This letter presents the results of the
observations issued from the CALAS project (a CAmera for the LArge
scales of the Sun; see \citealt{CALAS05}), collecting a sequence of
high resolution wide field images, which have allowed us to capture the
evolution of a hundred supergranules at the disc centre during 7.5~h. We
can thus give new constraints on the dynamics of the solar surface in
the supergranulation range. After a brief description of the data set
(Sect. 2), we show the spectral side of the supergranular flow (Sect.
3) as well as the probability distribution functions of the divergences
(Sect. 4); first conclusions follow.

\begin{figure*}
\centerline{\includegraphics[width=17cm]{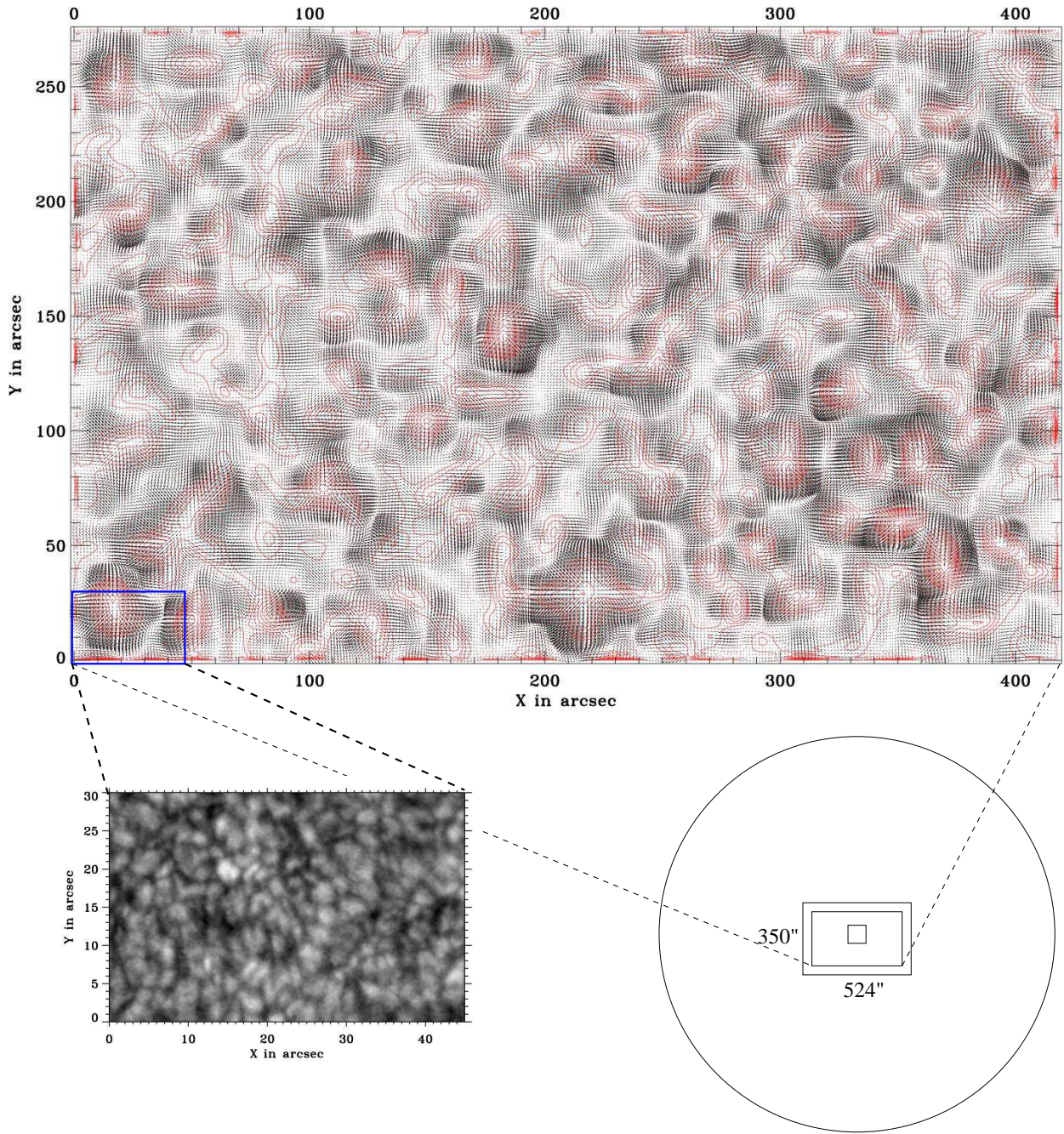}}
\caption[]{Bottom right: the CALAS field of view on the Sun; the large
rectangle indicates the size of the images, the smaller one shows the field
where the velocities could be computed, while the small square shows
the field of view of the SOT instrument on the Hinode satellite when
tracking granules. Top: the supergranulation velocity field with the
divergence contours superimposed (scales shorter than 8~Mm have been
filtered out). A time-window of 150~min was used. Bottom left: a zoom
on granulation showing the relative sizes of granules and supergranules.}

\label{calas_fields}
\end{figure*}

\section{Observational techniques and data reduction}

\subsection{Data set}
On 13 March 2007, we observed the Sun at disc centre during 7.5~h using
the Lunette Jean Rösch at Pic du Midi, a 50~cm-refractor. Images
were taken at $\lambda=575\pm5$nm, with a 14Mpixel CMOS-camera
(4560$\times$3048 pixels), with 0.115 arcsec/pixel, thus covering of
524$\times$350 arcsec$^2$ (see Fig.~\ref{calas_fields}).  10811 images
were obtained with a regular cadence of one every 2.5s. Two
independent series of $\sim1400$ images were then extracted. They sample
the same solar signal with a period of 20s but are noised differently by
the Earth atmospheric perturbations. The comparison between the outputs
of both series allows us to evaluate the influence of the seeing and
test the robustness of the results with respect to this noise.

Because of tracking difficulties, the common field of each series
was reduced to $\sim$400$\times$300 arcsec$^2$, thus covering a surface of
290$\times$216~Mm$^2$ on the Sun (Fig.~\ref{calas_fields}).

After recentering, subimages were $k-\omega$ filtered (with
a threshold of 7~km/s), so as to remove, as much as possible, Earth
atmospheric distortion, which is the main source of noise for velocity
measurements (\citealt{TRMR07}).

\vspace*{-5mm}
\subsection{Velocity fields}

\begin{figure}
\centerline{\includegraphics[width=8cm]{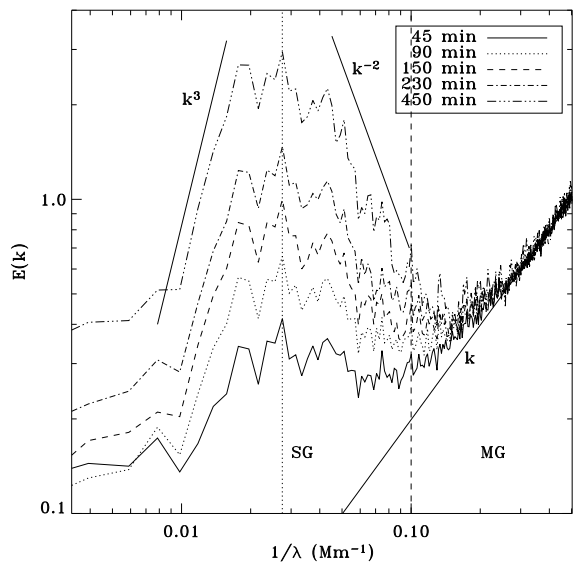}}
\caption[]{Kinetic energy spectra obtained for various time windows. The
vertical dotted line indicates the position of the peak at 36.4~Mm. The
vertical dashed line emphasizes the 10~Mm scale, usually taken as the
upper limit of mesogranular scale. Two power laws are shown on each side
of the peak, as well as the one of small-scale noise.}
\label{spectres}
\end{figure}

Horizontal velocity fields have been obtained using the CST granule tracking
algorithm (\citealt{RRMV99}; \citealt{RRRD07}). As shown in \cite{RRLNS01}, granules' motions
trace large-scale velocity fields when the scale is larger than 2.5~Mm.
Hence, we sampled the velocity field with a bin of 12 pixels ($\sim$1~Mm). 

The velocities are obtained by tracking the granules during a
given time window. Thus, we have access to average velocity components,
namely $\bv_x(i,j)$ and $\bv_y(i,j)$, where the overline refers
to the time averaging imposed by the time window.  This averaging
improves the signal-to-noise ratio, but naturally decreases the time
resolution. Typically, the shortest time window that can be used is
30~min.

Although the field is large, projection effects of the spherical Sun are
still of weak influence. At most, in the field corners, the correction
on the velocity would be less than 4\%, which is much less than the noise.

We show in Fig.~\ref{calas_fields} an example of these velocity
fields. The small-scale components of the flow (with scales below 8~Mm)
were filtered out using Daubechies wavelets (see \citealt{RRRD07}).
Figure~\ref{calas_fields} shows that robust steady supergranules live
among a wide variety of flow structures illustrating the turbulent nature
of these scales and their wide spectral range.

\vspace*{-0.5cm}
\section{The kinetic energy spectrum}

\begin{figure}
\centerline{\includegraphics[width=8cm]{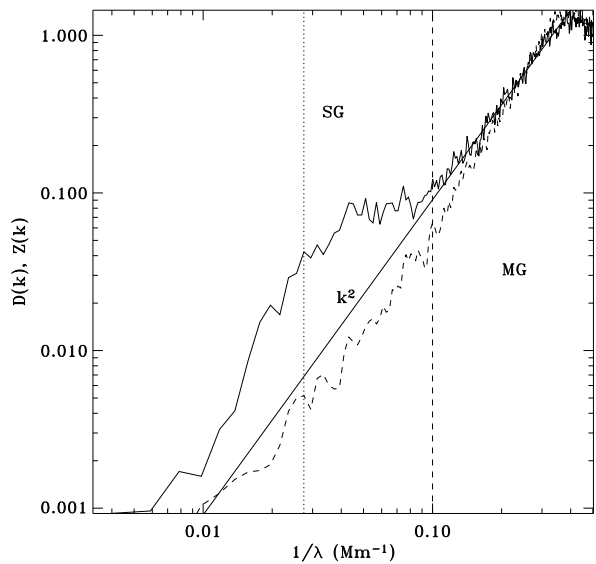}}
\caption{Same as in Fig.~\ref{spectres} but for the horizontal divergence
of the flow, $\partial_xv_x+\partial_yv_y$ (solid line), and for the
vertical vorticity, $\partial_xv_y-\partial_yv_x$ (dashed line); the
supergranulation peak is clearly visible at $\lambda=36$~Mm in the
divergence spectrum but absent in the vorticity one. A $k^2$-power law
is given for comparison.}\label{divrot}
\end{figure}

A convenient way to view the dynamics of a flow is to examine the
spectral content of the velocity field. We therefore computed the spectral
density of kinetic energy $E(k)$ associated with the horizontal flows that we
can measure. It is such that

\[ \demi<\bv^2>=\int_0^\infty E(k)dk\; .\]
$E(k)$, which is computed in the same way as in \cite{RRMR00}, is
displayed in Fig.~\ref{spectres} for the flows determined with various
time-windows. The short time-windows of 45~min give us 10 independent
spectra, which are averaged together, while the whole series 450~min
time-window gives us only one spectrum. Note that no spatial filtering
was applied to the velocity field. We see that $E(k)\propto k$ at small
scale, which is the signature of decorrelated random noise.

These spectra clearly show the emergence of the spectral range of the
supergranulation as the length of the time-window increases. Let us
point out the remarkable stability of the wavenumber of the spectral
peak when the time averaging is changed. This demonstrates
that supergranulation is a genuine velocity field at the
Sun's surface. It is not the consequence of time-averaging the small,
fast turning-over small scales. Of course, if the averaging interval is
long enough (i.e. of the order of the turn-over time scale of
supergranulation), the spectral peak will move to ever larger scales
before disappearing.  The maximum of the spectral density is at a
wavelength of 36~Mm. The FWHM of the peak indicates that supergranulation
occupies the range of scales of $[20,75]$~Mm.

The present mean value of the diameter of supergranules is slightly higher
than the previous determinations.  For instance, \cite{MTRR07} find a mean
diameter of $31.4$~Mm with a technique based on the segmentation of the
divergence field derived from velocities issued from a local correlation
technique applied to SOHO/MDI white light images. \cite{DBDK04} also
use a divergence field but derived from time-distance helioseismology;
they find a mean size of 27~Mm, significantly smaller than ours.

Both of these results are based on the divergence field and its
segmentation. The histogram of sizes is then used to determine the mean
diameter of a supergranule. Such a technique necessarily underestimates the
actual scale of supergranular flows as only part of it (the positive
divergences) is used. Our spectra, which directly result from the
measured horizontal velocities, incorporate all the components of the
flow at supergranulation scale, and thus better reflect the
dynamical state. Note that a blind use of the divergence spectrum in
Fig.~\ref{divrot} would point to a scale of $\sim23$~Mm.

In Fig.~\ref{spectres} we also indicate the best-fit power laws, which
mimic the sides of the supergranulation peak. We find that $E(k) \sim
k^3$ on the large-scale side and $E(k) \sim k^{-2}$ on the small-scale
one. This latter power law is steeper than the $-5/3$ Kolmogorov one
and may be an effect of density stratification.

To complete  the spectral picture, we also show, in Fig.~\ref{divrot},
the spectra of the horizontal divergence of the velocity field and the
vertical vorticity. Beyond the $k^2$-dependence of the spectral densities,
which is a consequence of the derivative of uncorrelated noise, we can see
that the divergence spectrum clearly shows the supergranulation peak, while
the vertical vorticity shows no signal in this range. The vorticity does
show some weak signal however, but in the mesogranulation range, below 10~Mm.

\begin{figure}
\includegraphics[width=7.3cm,angle=0]{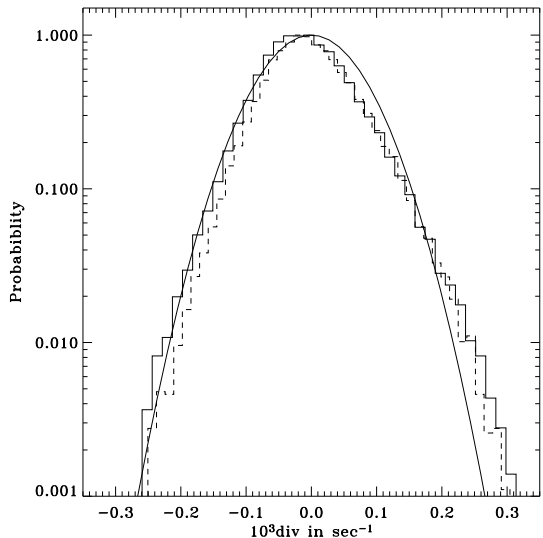}
\caption[]{Histograms of the divergences; solid and dashed lines have the
same solar signal but a different noise from the Earth atmosphere. Scales
below 8~Mm have been filtered out. A Gaussian distribution with the same
standard deviation is overplotted for comparison.} \label{histo}
\end{figure}

\vspace*{-0.5cm}
\section{Probability density functions}

Another way to look at the velocity fields is to consider the probability
density functions. Such distributions have been computed for the
velocity field, its divergence and its curl. We show in Fig.~\ref{histo}
the distribution of the divergence field when scales shorter than 8~Mm
have been filtered out of the velocity field.  This distribution clearly
shows a positive wing larger than the Gaussian one, associated with a
skewness of 0.38. We also note that the noise of the Earth atmosphere is weak
enough, and does not perturb the result. This trend to an exponential
distribution is a signature of intermittency (e.g. \citealt{Frisch95}).  This
result, with the same asymmetry between positive and negative values,
was also observed by \cite{MTRR07} with a completely different set of
data and method. Thus, the intermittency of supergranulation seems to
be a robust property. Exponential wings are usually less visible on the
velocity (e.g. \citealt{VM91}), and, indeed, are barely noticeable in our
data. As far as vorticity is concerned, the noise is unfortunately too
high to give convincing measurements.

\vspace*{-0.5cm}
\section{Conclusions}

For the first time, it has been possible to follow the motion of
well-resolved granules in a large field of view. The spectral peak of
supergranulation has thus been determined in a very direct
manner. According to our data, supergranulation has the most energetic
motions at a scale of 36~Mm and encompasses all the scales ranging from 20
to 75~Mm, as shown by the FWHM of the peak. Except for its amplitude, this
peak is not sensitive to the time window used to measure the granules'
motions. The signal also clearly appears in the divergence spectral
density, but not in the vorticity; it is likely that vorticity at supergranulation
scale near the Sun's equator is much weaker and does not emerge from
the noise.

Finally, our data confirm the fact that supergranulation has a noticeable
degree of intermittency, clearly appearing in the distribution of
positive divergence values.

As far as the origin of supergranulation is concerned, the scenarios
mentioned in the introduction can be tested with these data, either in the
real space with velocity fields like the one shown in Fig.~\ref{calas_fields}
or in the spectral space with the given spectra.

Further work on the observational side will focus on the determination
of the third component of the velocity field, the increase of the field
size and the reduction of the noise, so as to further constrain the
dynamics of scales from the granulation one to the 100~Mm one.

\vspace*{-0.2cm}
\begin{acknowledgements}
The CALAS project has been financially supported by the French ministery
of education (ACI), by the Programme National Soleil-Terre of CNRS and
the Observatoire Midi-Pyrénées. We are also very grateful to the ``Groupe
d'Intrumentation des Grands Télescopes (GIGT)" of the laboratory,
for their technical help at various phase of the project, especially
to Sébastien Baratchart and Elodie Bourrec. We also wish to
thank René Dorignac for his efficient support in mechanical realizations
and Philippe Saby for his help in sorting out the right computing
hardware. SR wishes to thank the CNRS for its support during his PhD
thesis which much contributed to the project.
\end{acknowledgements}

\vspace*{-0.8cm}

\bibliographystyle{aa}

\end{document}